\documentclass[aps,prx,twocolumn,
			   groupedaddress,superscriptaddress,
			   amsfonts,amssymb,amsmath,
			   citeautoscript,longbibliography,
			   a4paper
			   ]{revtex4-2}

\usepackage[utf8]{inputenc}
\usepackage[english]{babel}

\usepackage{microtype} 
\usepackage{xspace} 

\usepackage{txfonts}  
\usepackage{txfontsb} 

\usepackage{bm} 

\usepackage{xcolor}
\usepackage[]{graphicx} 
\graphicspath{{figs/}}

\usepackage[]{booktabs}
\usepackage{array}
\usepackage{layouts}
\usepackage{multirow}

\usepackage{enumerate}
\usepackage[inline]{enumitem}

\usepackage{hyperref}
\hypersetup{colorlinks,
	linkcolor={blue!75!black!80!yellow},
	citecolor={blue!75!black!80!yellow},
	urlcolor={blue!75!black!80!yellow}
}

\usepackage[capitalize,nameinlink]{cleveref}

\crefname{subequations}{Eqs.}{Eqs.} 
\Crefname{subequations}{Eqs.}{Eqs.}
\crefformat{subequations}{#2Eqs.~(#1)#3}
\Crefformat{subequations}{#2Eqs.~(#1)#3}
\crefname{page}{p.}{p.} 

\usepackage{placeins}

\usepackage{siunitx}
\DeclareSIUnit[number-unit-product = ]\percent{\char`\%} 

\usepackage[centering,hmargin=16mm,tmargin=30mm,bmargin=26mm]{geometry}

\usepackage{soul}

\newcommand{\rv}{\mathbf{r}}

\newcommand{\kv}{\mathbf{k}}
\newcommand{\Ev}{\mathbf{E}}
\newcommand{\Dv}{\mathbf{D}}

\newcommand{\Bv}{\mathbf{B}}

\newcommand{\ev}{\mathbf{e}}
\newcommand{\nablav}{\boldsymbol{\nabla}}

\newcommand{\nv}{\mathbf{n}}
\newcommand{\hv}{\mathbf{h}}

\newcommand{\e}{\mathrm{e}}

\newcommand{\iu}{\mathrm{i}}

\newcommand{\ie}{i.e.,\@\xspace} 

\newcommand{\eg}{e.g.,\@\xspace}

\newcommand{\appropto}{\mathrel{\vcenter{
			\offinterlineskip\halign{\hfil$##$\cr
				\propto\cr\noalign{\kern.2pt}\sim\cr\noalign{\kern-2.5pt}}}}}

\newcommand{\Mod}[1]{\ (\mathrm{mod}\ #1)}

\newcommand{\BS}{\{BS\}\xspace}

\newcommand{\suppsec}{Supplemental Section\xspace}

\newcommand{\muT}{\mu^{\text{T}}}
\newcommand{\muL}{\mu^{\text{L}}}
\newcommand{\muLT}{\mu^{\text{L+T}}}
\newcommand{\muTbtr}{\tilde{\mu}^{\text{T}}}

\DeclareFontFamily{U}{mathb}{}
\DeclareFontShape{U}{mathb}{m}{n}{
  <-5.5> mathb5
  <5.5-6.5> mathb6
  <6.5-7.5> mathb7
  <7.5-8.5> mathb8
  <8.5-9.5> mathb9
  <9.5-11.5> mathb10
  <11.5-> mathbb12
}{}
\DeclareRobustCommand{\smallsquare}{\mathbin{\text{\usefont{U}{mathb}{m}{n}\symbol{"0D}}}}

\DeclareFontFamily{U}{mathx}{\hyphenchar\font45}
\DeclareFontShape{U}{mathx}{m}{n}{<5> <6> <7> <8> <9> <10>
                                  <10.95> <12> <14.4> <17.28> <20.74> <24.88>
                                  mathx10}{}
\DeclareSymbolFont{mathx}{U}{mathx}{m}{n}
\DeclareFontSubstitution{U}{mathx}{m}{n}
\DeclareMathAccent{\widebar}{0}{mathx}{"73}

\makeatletter
\newcommand{\raisemath}[1]{\mathpalette{\raisem@th{#1}}}
\newcommand{\raisem@th}[3]{\raisebox{#1}{$#2#3$}}
\makeatother

\renewcommand{\paragraph}[1]{\vskip 1ex\noindent\textbf{#1.}~}

\usepackage[eulergreek]{sansmath}
\makeatletter
\renewcommand\@make@capt@title[2]{%
    \@ifx@empty\float@link{\@firstofone}{\expandafter\href\expandafter{\float@link}}%
    \sisetup{math-sf=\textsf}%
    \sansmath\sffamily\textbf{#1\@caption@fignum@sep}#2 
}%

\makeatother

\usepackage{xr}
\makeatletter
\newcommand*{\addFileDependency}[1]{
  \typeout{(#1)}
  \@addtofilelist{#1}
  \IfFileExists{#1}{}{\typeout{No file #1.}}
}
\makeatother

\newcommand*{\myexternaldocument}[1]{%
    \externaldocument{#1}%
    \addFileDependency{#1.tex}%
    \addFileDependency{#1.aux}%
}

\usepackage[final]{pdfpages}
\makeatletter 
\AtBeginDocument{\let\LS@rot\@undefined}
\makeatother
\usepackage{pgffor} 

\myexternaldocument{supp/supp}

\newcommand{\captionbullet}[1]{\textbf{#1}}
\graphicspath{{figs/}}

\begin{document}
\title{Location and topology of the fundamental gap in photonic crystals}

\def\mitaffil{Department of Physics, Massachusetts Institute of Technology, Cambridge, Massachusetts, USA}
\def\hkustaffil{Department of Physics, Hong Kong University of Science and Technology, Clear Water Bay, Hong Kong, China}
\author{Thomas~Christensen}
\email{tchr@mit.edu}
\affiliation{\mitaffil}
\author{Hoi~Chun~Po}
\affiliation{\mitaffil}
\affiliation{\hkustaffil}
\author{John D. Joannopoulos}
\affiliation{\mitaffil}
\author{Marin Solja\v{c}i\'{c}}
\affiliation{\mitaffil}

\begin{abstract}    
    The fundamental, or first, band gap is of unmatched importance in the study of photonic crystals.
    Here, we address precisely where this gap can be opened in the band structure of three-dimensional photonic crystals.
    Although strongly constrained by symmetry, this problem cannot be addressed directly with conventional band-symmetry analysis due to the existence of a photonic polarization vortex at zero frequency.
    We develop an approach for overcoming the associated symmetry singularity by incorporating fictitious, auxiliary longitudinal modes.
    Our strategy also enables us to extend recent developments in symmetry-based topological analysis to the fundamental gap of three-dimensional photonic crystals.
    Exploiting this, we systematically study the topology of the minimal fundamental gaps.
    This reveals the existence of topological gap-obstructions that push the fundamental gap higher than what a conventional analysis would suggest.
    Our work demonstrates that topology can play a crucial role in the opening of the fundamental photonic gap and informs future theoretical and experimental searches for conventional and topological band gaps in three-dimensional photonic crystals.
\end{abstract}
\maketitle

\section{Introduction}

The pursuit of photonic band gaps has been a key driving force in the field of photonic crystals (PhCs)~\cite{Joannopoulos:2008, Sakoda:2004}, from Rayleigh's earliest treatments of one-dimensional PhCs in 1887~\cite{Rayleigh:1887, Rayleigh:1888}, to Yablonovitch's~\cite{Yablonovitch:1987} and John's~\cite{John:1987} three-dimensional (3D) generalizations a century later, and continuing to this day~\cite{Maldovan:2003, CersonskyGlotzer:2021}.
The recent incorporation of ideas from topological band theory~\cite{Hasan:2010, Bernevig:2013, Bansil:2016} to photonics~\cite{Haldane:2008, Lu:2014, Ozawa:2019} has reinvigorated this fascination by highlighting that PhC bands---and the gaps between them---can possess robust topological properties.
Armed with recent insights from topological band theory, we address one of the fundamental problems in the study of 3D PhCs: where is the lowest photonic band gap in the band structure and what is its topology?

\begin{figure}[!htb]
	\centering
	\includegraphics[scale=1]{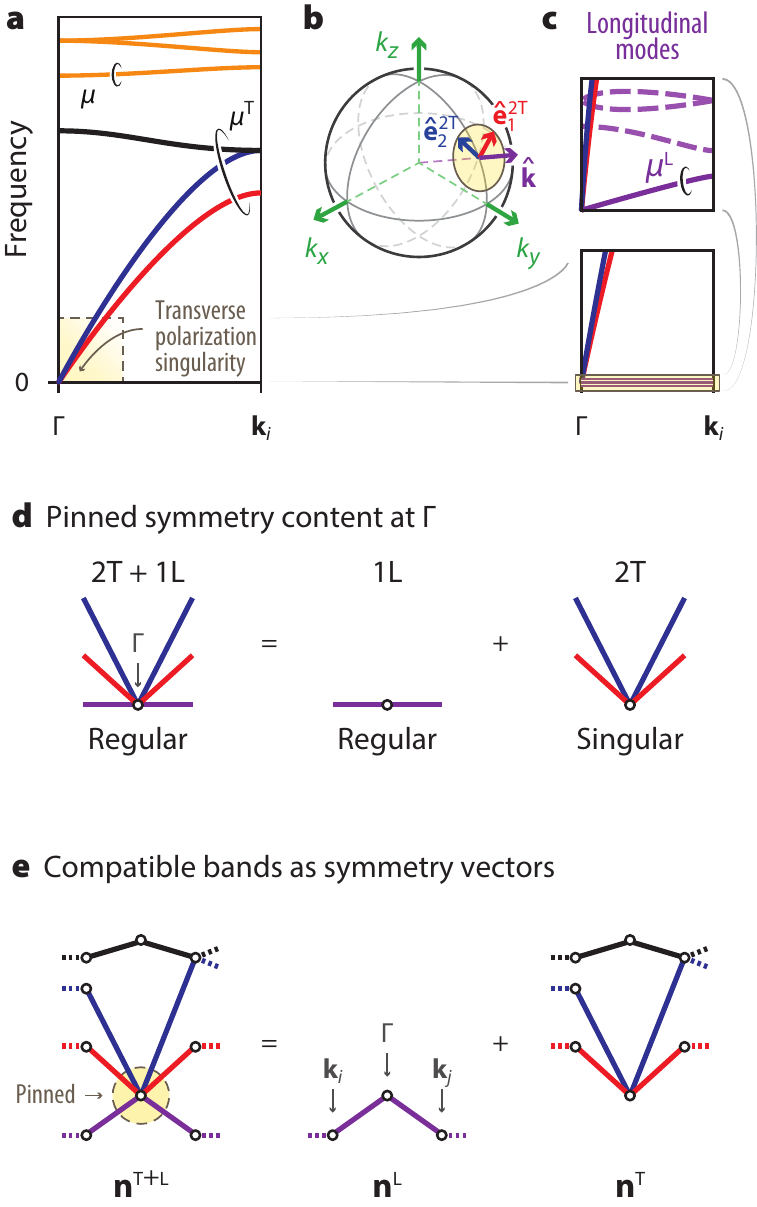}
	\caption{%
	    \textbf{PhC band connectivity below the fundamental gap.}
	    \captionbullet{a}~The two lowest-frequency ``light-like'' PhC bands (blue and red) are pinned to $\omega=0$ at the $\mathsf{\Gamma}$-point.
        The minimum connectivity $\muT$ of these bands, \ie, below the fundamental gap, is not generally equal to the minimum connectivity $\mu$ of higher bands (orange).
	    \captionbullet{b}~The polarization vectors $\hat{\mathbf{e}}_{1,2}^{\text{2T}}$ of the transverse modes near $\omega=0$ span a space (yellow disk) that varies with wave vector orientation $\hat{\kv}$, rendering their associated symmetry content at $\mathsf{\Gamma}$ singular.
	    \captionbullet{c}~In addition to physical, transverse PhC modes, the Maxwell equations also admit a set of nonphysical, longitudinal modes (purple).
        \captionbullet{d}~The singularity can be regularized by adding a longitudinal mode (1L) to the two transverse modes (2T) at $\omega=0$.
        \captionbullet{e}~Realizable band configurations connected to $\omega=0$ correspond to compatibility-allowed symmetry vectors $\nv^{\text{T+L}}$ of the combined transverse and longitudinal band set, itself a sum of longitudinal and transverse contributions $\nv^{\text{L}}$ and $\nv^{\text{T}}$.
		\label{fig:1}
	}
\end{figure}

In more precise terms, we ask how many bands are required, at minimum, by spatial and time-reversal (TR) symmetry below the first photonic gap (\cref{fig:1}a), \ie where it can be opened.
The first, or fundamental, gap is special---and of particular interest---for two reasons:
(1)~the first photonic gap is usually the largest and most easily engineered~%
    \footnote{The perturbation of a dielectric profile $\boldsymbol{\varepsilonup}(\rv)$ on an empty-lattice band of frequency $\omega_{n\kv} = c|\mathbf{G}+\kv|_n$ is dominated by the corresponding Fourier components of $\boldsymbol{\varepsilonup}(\rv)$, \ie of those $\boldsymbol{\varepsilonup}_{\mathbf{G}'\mathbf{G}'}$ with $|\mathbf{G}' + \kv|=|\mathbf{G}+\kv|_n$.
    Fabrication constraints typically makes $|\boldsymbol{\varepsilonup}_{\mathbf{G}'\mathbf{G}'}|$ decrease rapidly with $|\mathbf{G}'|$, corresponding to reduced control of higher bands.}
and
(2)~the bands below the first gap in a PhC are unlike all other bosonic quasiparticle bands, because they connect to a polarization singularity at zero frequency ($\omega=0$) and zero momentum ($\kv = \boldsymbol{0}$, \ie $\Gamma$).
This singularity arises due to the transverse polarization of photons (\cref{fig:1}b) and has profound implications for the first gap: in general, the number of bands required below the first gap is different from the that required between higher-lying gaps, since the latter are not affected by the singularity~\cite{Watanabe:2018, Michel:1999, Watanabe:2016}.
The singularity additionally renders the band symmetry at $\Gamma$ ill-defined, ostensibly preventing application of symmetry-based topological analysis to the most important gap of 3D PhCs~\cite{Bradlyn:2019}.

A systematic study of this minimum-connectivity problem for PhCs was recently initiated by \citet{Watanabe:2018}, who derived various lower bounds for the number of connected bands using sub- and supergroup relations. 
Given a specific PhC, however, it is unclear in general if these lower bounds have already made maximal use of the present spatial symmetries.
In addition, the topology of the bands below the first gap remains unaddressed.

Here, we develop new tools that enable us to establish the exact minimum connectivity and topology below the first gap for PhCs in each of the 230 possible symmetry settings, \ie space groups.
Our approach extends recent symmetry-based tools for topological band analysis in condensed-matter systems~\cite{KruthoffSlager:2017, Po:2017, Bradlyn:2017} to the fundamental gap of 3D PhCs, overcoming the apparent barrier raised by the singular $\Gamma$-point symmetry.
Surprisingly, we find that complete answers to the photonic minimum-connectivity question cannot be obtained without topological considerations.
Specifically, we find six space groups whose minimum-connectivity bands are all topologically nontrivial, entailing a topological obstruction to the opening of the symmetry-allowed gap. 
This effect, which we term ``$\Gamma$-enforced topology,'' resembles filling-enforced topology~\cite{Po:2016, Wang:2020} but arises here as a direct consequence of the zero-frequency $\Gamma$-point singularity.
For PhCs with $\Gamma$-enforced topology, observing a minimum-connectivity fundamental gap along the high-symmetry lines implies that the gap must close at nodal lines in the interior of the Brillouin zone (BZ). 
This pushes the true fundamental gap to the higher-energy part of the spectrum---an insight from topological band theory that cannot be inferred from conventional symmetry analysis alone.
Finally, as the analogous minimum-connectivity problem for \emph{phonons} occurs as a sub-problem in our approach, we solve the phononic problem as well.

\section{Theoretical framework}

The two lowest-frequency transverse (2T) solutions of non-metallic PhCs (\ie with periodic and non-negative dielectric $\varepsilonup$ and magnetic $\muup$ response profiles) asymptotically realize the effective medium approximation at small momenta $|\kv|\rightarrow 0$ where they touch $\omega=0$ with the ``light-like'' linear dispersion $(\overline{\varepsilonup^{-1}\muup^{-1}})^{1/2}c|\kv|$ (\cref{fig:1}a).
Their displacement fields converge to plane-waves $\mathbf{D}^{\text{2T}}_{\sigma\kv}(\rv) \sim \hat{\ev}^{\text{2T}}_{\sigma\kv}\e^{\iu\kv\cdot\rv}$ ($\sigma = 1,2$) with mutually orthogonal polarization vectors $\hat{\ev}^{\text{2T}}_{\sigma\kv}$ polarized transversely to the wave vector $\kv$.
The resulting vortex-like polarization texture around the $\Gamma$-point (\cref{fig:1}b) renders the continuation of the 2T solutions to $|\kv| = 0$ ill-defined (it depends on $\hat{\kv}$) and is the salient feature that distinguishes photons from all other bosonic quasiparticles (\eg phonons, excitons, or magnons) in the present context.

The above considerations show that the light-like 2T modes behave as free photons (in an effective medium) at small momenta. 
As we go to higher momenta, however, the effects of the PhC becomes apparent.
The periodicity of the PhC leads to BZ folding and photonic band gaps arise when energetically intersecting bands anti-cross with each other.
Such anti-crossing is heavily constrained by the spatial symmetries of the PhC, since the hybridization matrix elements between bands with distinct symmetry characters vanish.
Importantly, symmetry eigenvalues of rotations and mirrors as well as their nonsymmorphic counterparts---screw and glides---are continuously defined over lines and planes in the BZ.
For the lowest-lying states these symmetry eigenvalues of the bands can be traced down to the singularity at the zero-frequency limit~\cite{Watanabe:2018}.
This leads to a dilemma in the symmetry analysis of the fundamental photonic band gap: while the symmetry characters of the lowest bands are dictated by those of the lowest lying light-like modes, their symmetry eigenvalues are ill-defined in the zero-momentum limit.

In the following, we outline our theoretical framework and strategy for overcoming the singularity problem. 
We first describe the fundamental relationship between band connectivity, compatibility relations, and band symmetries and introduce a scheme to generate all allowable band symmetries through stacking of a minimal set of intrinsically connected bands (\cref{subsec:compatibility}).
We then discuss the implications of the polarization vortex in the context of pinned symmetry and compatibility relations (\cref{subsec:singularity}) and introduce a regularization strategy featuring auxiliary apolar and longitudinal modes (\cref{subsec:transverse_sols}).
Following this overview, \cref{sec:results} considers the results obtained from the application of our theoretical framework.
The results for the minimum photonic connectivity below the first gap are given first (\cref{subsec:connectivity}).
Next, we introduce a scheme to evaluate the symmetry-identifiable topology of the otherwise singular photonic bands below the first gap and use it to determine the topology of all minimum-connectivity solutions (\cref{subsec:topology}).
Our analysis uncovers six space groups whose minimum-connectivity solutions are all topologically nontrivial and consequently display a topological obstruction to gap-opening.
We end by introducing two concrete PhCs that demonstrate this effect (\cref{subsec:nongaps}).

\subsection{Compatibility relations and band connectivity}\label{subsec:compatibility}
The presence of crystalline symmetries constrains the possible connections between energy bands across the BZ due to the existence of compatibility relations~\cite{Bouckaert:1936, Cornwell:1969, Kittel:1987, Kaxirax:2019}.
These relations express a familiar notion, namely the splitting of symmetry-protected degeneracies in finite systems, but translated to $\kv$-space.
In finite systems, such degeneracies can be split only by deforming the considered object to a state of lowered symmetry (\ie by lowering the point group symmetry).
In crystalline systems, however, the modal symmetry at any given $\kv$-point in the band structure is determined by the subset of the space group symmetry that additionally leaves the considered $\kv$-point invariant (\ie by the little group of $\kv$).
Compatibility relations express how modal degeneracy and symmetry is reduced or maintained as we move off an initial high-symmetry point towards a lower-symmetry line (or plane).
Since such lines can connect to other high-symmetry $\kv$-points---\eg starting at $\Gamma$ and moving towards $k_x$, one reaches the high-symmetry point $X$ at the BZ boundary---and since continuity requires modal symmetry to be invariant along lines of fixed symmetry, the compatibility relations at distinct high-symmetry points become coupled, effectively tying together multiple local relations into a global set of consistency constraints spanning the BZ.
These global constraints restrict how (and how many times) bands must connect and, crucially, when they can be gapped~\cite{Michel:1999, Michel:2001}.

For any given space group, the solutions to the aggregate set of compatibility constraints (along with a requirement that all symmetry data be non-negative) defines the set of all physically realizable band structures \BS.
Each element of \BS can be identified with a ``symmetry vector'' $\nv$ that enumerates the symmetry content of the included bands across all non-equivalent $\kv$-points in the BZ.
The elements of $\nv$ give the multiplicity $n_{\kv_i}^{\alpha}$ of the $\alpha$th small irreducible representation (irrep) $D_{\kv_i}^\alpha$ in the little group $G_{\kv_i}$ at $\kv_i$~%
    \footnote{We use the data tables of ISO-IR~ to construct small irreps~\cite{Stokes:2013, ISO-IR}, accessed with the tooling developed in Ref.~\citenum{Crystalline.jl}.
    The associated irrep labels follow the CDML notation~\cite{CDML:1979}, consistently with \eg the Bilbao Crystallographic Server~\cite{Elcoro:2017}.}%
, such that 
\begin{equation}\label{eq:symvec}
    \nv
    \equiv
    \big[ n_{\kv_1}^\alpha, n_{\kv_1}^\beta, \ldots, n_{\kv_i}^{\alpha'}, n_{\kv_i}^{\beta'}, \ldots, \mu \big]^{\mathrm{T}},
\end{equation}
with the number of included bands $\mu$ incorporated as well.
We denote by $\nv_{\kv_i} \equiv [n_{\kv_i}^\alpha, n_{\kv_i}^\beta, \ldots]^{\mathrm{T}}$ the projection of $\nv$ to its symmetry at the $\kv_i$ point.
To obtain $\nv_{\kv_i}$ for a band grouping $\{n'\}$, we first compute the symmetry eigenvalues $x_{n\kv_i}(g) \equiv \langle \Ev_{n\kv_i} | g\Dv_{n\kv_i} \rangle$ for each operation $g\in G_{\kv_i} = \{g_1, \ldots, g_{|G_{\kv_i}|}\}$ and each band $n\in\{n'\}$; 
next, we aggregate eigenvalues in the character vector  $\mathbf{x}_{\kv_i} \equiv  \sum_n \big[x_{\kv_i}(g_1), \ldots, x_{\kv_i}(g_{|G_{\kv_i}|})\big]^{\text{T}}$;
and finally, solve $\boldsymbol{\chi}_{\kv_i}\nv_{\kv_i} = \mathbf{x}_{\kv_i}$ for $\nv_{\kv_i}$ with $\boldsymbol{\chi}_{\kv_i}$ denoting the character table of $G_{\kv_i}$ with characters $\chi_{\kv_i}^{\alpha}(g) \equiv \mathop{\mathrm{Tr}}D_{\kv_i}^\alpha(g)$ operator-indexed ($g$) along rows and irrep-indexed ($\alpha$) along columns.
The $\Ev$- and $\Dv$-fields transform as vector fields, \ie $g\Dv_{n\kv}(\rv) = (g\Dv_{n\kv})(g^{-1}\rv)$~\cite{Sakoda:2004}.

The structure of \BS has been explored using both graph theory~\cite{Bradlyn:2017, Vergniory:2017, Bradlyn:2018} and linear algebra~\cite{KruthoffSlager:2017, Po:2017, Elcoro:2020}.
Here, inspired by Ref.~\citenum{Song:2020}, we pursue a different approach which allows assembling \BS from a set of minimal and intrinsically connected bands.
First, we note that elements of \BS are equipped with a composition operation, namely ``stacking'' of bands (addition of symmetry vectors), but not with an analogous inverse operation (subtraction of symmetry vectors need not be physical, \ie \BS is not complete under subtraction).
This describes the algebraic structure of a monoid: a group lacking inverse operations.
More precisely, \BS is a positive affine monoid (it is a submonoid of a free abelian group~\cite{Po:2017} and bounded by a pointed polyhedral cone) and is therefore equipped with a unique, minimal basis $\{\mathbf{h}_i\}$---a Hilbert basis~\cite{Bruns:2009}---whose non-negative integer combinations generate \BS:
\begin{equation}\label{eq:hilbert}
    \{\text{BS}\} 
    =
    \biggl\{\sum_i c_i \mathbf{h}_i \Bigm\vert c_i\in0,1,\ldots\biggr\}.
\end{equation}
As we show in \suppsec~\ref{supp-sec:A}, the basis $\{\mathbf{h}_i\}$ can be derived from a related basis, namely the elementary band representations (EBRs) of topological quantum chemistry~\cite{Bradlyn:2017} whose ``stacking'' generates the set of all topologically trivial (or ``atomic'') insulators
(in summary, we define \BS as the intersection of a lattice and a polyhedral cone and obtain the associated Hilbert basis using the Normaliz software~\cite{NormalizSoftware}).
Crucially, the basis vectors $\mathbf{h}_i$ necessarily describe connected bands; if they were not, they would not form a \emph{minimal} basis for \BS.
Conceptually, the elements of $\{\mathbf{h}_i\}$ are the indivisible units whose stacking yield all separable band structures.

\subsection{Pinned symmetry content at $\Gamma$}\label{subsec:singularity}
At first glance, the existence and properties of a Hilbert basis for \BS would appear to solve the question of band connectivity entirely.
Indeed, if we denote the connectivities associated with a Hilbert basis $\{\mathbf{h}_i\}$ by $\{\mu_i\}$, the minimum realizable connectivity below any gap between \emph{regular} bands (\eg electrons) is just $\min\{\mu_i\}$.
As noted earlier, however, photonic bands below the first gap are \emph{not regular}, because of the ill-definiteness of the 2T solutions at $\Gamma$. 
This ill-definiteness extends to the symmetry vector for 2T-connected bands, presenting a clear obstacle.

To overcome this, we first describe how a partial, effective assignment of the 2T $\Gamma$-point symmetry can be constructed.
Specifically, we can treat the 2T $\Gamma$-point symmetry  as a surrogate for compatibility constraints imposed by line and plane little groups that intersect $\Gamma$.
These ``interior'' little groups can include proper rotations and screws $r_\theta$ (of angle $\theta$) as well as mirrors and glides, $m$.
Their symmetry eigenvalues can be evaluated directly from our knowledge of the asymptotic 2T fields at small $|\kv|$ and then continued to $\Gamma$, giving $\e^{\pm\iu \theta}$ and $\pm1$, respectively, as noted by \citet{Watanabe:2018}.
The associated characters, \ie sum of symmetry eigenvalues, are $x_{\Gamma}^{\text{2T}}(r_{\theta}) \equiv 2\cos\theta$ and $x_{\Gamma}^{\text{2T}}(m) = 0$.

Unlike rotations and mirrors, improper rotations and inversions can only be symmetries at $\Gamma$ or at the BZ boundary, and so are not similarly continuable to $\Gamma$.
The surrogate $\Gamma$-point irrep is therefore underdetermined for space groups with (roto-)inversions.
Moreover, even for the 113 space groups without (roto-)inversions, the surrogate 2T irrep can be singular.
As an example, space group 16 (P222) consists of operations $\{1, 2_{001}, 2_{010}, 2_{100}\}$ and has the 2T $\Gamma$-point character vector $\mathbf{x}_\Gamma^{\text{2T}} \equiv [x_\Gamma^{\text{2T}}(1), x_\Gamma^{\text{2T}}(2_{001}), x_\Gamma^{\text{2T}}(2_{010}), x_\Gamma^{\text{2T}}(2_{100})]^{\text{T}} = [2, -2, -2, -2]^{\text{T}}$ (in CDML notation~\cite{CDML:1979}).
The associated $\Gamma$-projected symmetry vector is $\nv_{\Gamma}^{\text{2T}} =  -\Gamma_1+\Gamma_2+\Gamma_3+\Gamma_4$.
Notably, this includes a \emph{subtracted} irrep $\Gamma_1$, which prevents an expansion in the Hilbert basis $\{\mathbf{h}_i\}$ (which is strictly non-negative): the 2T symmetry at $\Gamma$ is singular.

We now introduce new techniques to regularize this singularity.
To that end, we first observe that the Maxwell equations admit not only transverse (divergence-free) solutions but also longitudinal (curl-free) solutions which, however, violate the transversality condition $\nablav\cdot\Dv = \nablav\cdot\Bv = 0$ (unless $\varepsilonup$ or $\muup$ vanishes).
In local media, the longitudinal solutions are completely degenerate with eigenfrequencies $\omega_{n\kv}=0$.
It is useful to imagine lifting their dispersion to a more conventional band structure (which, physically, can be achieved \eg by including nonlocality and a weak Drude term in the material response) as illustrated in \cref{fig:1}c.
Even in this ``lifted'' picture, a single longitudinal solution (1L) will always connect to $\omega=0$ at $\Gamma$ with an asymptotic plane-wave-like field profile $\Dv^{\text{1L}}_{\mathbf{k}}(\rv) \sim \hat{\kv}\e^{\iu\kv\cdot\rv}$.
Its $\Gamma$-continuable characters transform trivially, \ie $x^{\text{1L}}_\Gamma(r_\theta) = x^{\text{1L}}_\Gamma(m) = 1$, and we are free to also \emph{choose} the characters of any ``unpinned'' roto-inversions to transform trivially, in which case the 1L band always transforms trivially as a whole at $\Gamma$ (\ie as $\Gamma_1$ or $\Gamma_1^+$).

More importantly, the symmetry content of the ``apolar'' sum of 1L and 2T bands, \ie $\nv_\Gamma^{\text{1L}+\text{2T}} = \nv_\Gamma^{\text{1L}} + \nv_\Gamma^{\text{2T}}$, is well-defined and regular (\cref{fig:1}d) because the space spanned by their combination is invariant to $\hat{\kv}$ in the $|\kv|\rightarrow 0$ limit.
The associated characters are uniquely determined for both proper ($+$) and improper ($-$) rotations $g$, equaling $x^{\text{1L}+\text{2T}}_{\Gamma}(g) = \pm2\cos\theta \pm 1$.
Considering again space group 16, we have $\nv_\Gamma^{\text{1L}+\text{2T}} = \Gamma_2+\Gamma_3+\Gamma_4$ and $\nv_\Gamma^{\text{1L}} = \Gamma_1$, revealing that the negative $\Gamma_1$ irrep in $\nv_\Gamma^{\text{2T}}$ simply represents a subtraction of the 1L mode.
We therefore generally define the surrogate 2T symmetry at $\Gamma$ as $\nv_\Gamma^{\text{2T}} = \nv_\Gamma^{\text{1L}+\text{2T}} - \nv_\Gamma^{\text{1L}}$, \ie as the subtraction of two regular representations.

\subsection{Transverse band solutions}\label{subsec:transverse_sols}
The $\Gamma$ symmetry content at $\omega=0$, \ie $\nv_\Gamma^{\text{1L}}$, $\nv_\Gamma^{\text{2T}}$, and $\nv_\Gamma^{\text{1L}+\text{2T}}$, imposes constraints%
\begin{equation}\label{eq:gamma_constraints}
    \nv_\Gamma^{\text{L}} \geq \nv_\Gamma^{\text{1L}},\qquad
    \nv_\Gamma^{\text{T}} \geq \nv_\Gamma^{\text{2T}},\qquad
    \nv_\Gamma^{\text{L}+\text{T}} \geq \nv_\Gamma^{\text{1L+2T}},
\end{equation}
on the symmetry vectors of the longitudinal ($\nv^{\text{L}}$), transverse ($\nv^{\text{T}}$), and apolar ($\nv^{\text{T}+\text{L}}$) band solutions connected to $\omega=0$ (\cref{fig:1}e).
Since $\nv_\Gamma^{\text{1L}}$ and $\nv_\Gamma^{\text{1L}+\text{2T}}$ are regular, the auxiliary solutions $\nv^{\text{L}}$ and $\nv^{\text{L}+\text{T}}$ can always be expanded in the Hilbert basis $\{\mathbf{h}_i\}$.
The same is not generally possible for the transverse solutions $\nv^{\text{T}}$, however, since the constraint $\nv_\Gamma^{\text{2T}}$ may be singular.
Candidate transverse solutions can instead be obtained as the subtraction of an apolar and a longitudinal solution, \ie as $\nv^{\text{T}} = \nv^{\text{T}+\text{L}} - \nv^{\text{L}}$.
Those candidates that respect the $\Gamma$ constraints of \cref{eq:gamma_constraints} and are regular at all other $\kv$-points, \ie have
\begin{equation}\label{eq:regularity_constraint}
    \nv_{\kv_i}^{\text{T}} \geq 0,
\end{equation}
for all $\kv_i\neq\Gamma$, correspond to physically realizable transverse bands connected to $\omega=0$.
If we denote by $\{\nv^{\text{L+T}}\}$ the set of all apolar solutions and by $\tilde{\nv}^{\text{L}}$ some longitudinal solution, each consistent with \cref{eq:gamma_constraints}, then all transverse solutions $\{\nv^{\text{T}}\}$ can be identified with the elements of the set $\{\nv^{\text{L+T}}\} - \tilde{\nv}^{\text{L}}$ that are consistent with \cref{eq:gamma_constraints,eq:regularity_constraint}.
The specific choice of $\tilde{\nv}^{\text{L}}$ is immaterial since the transverse and auxiliary longitudinal degrees of freedom are decoupled (except at $\omega=0$).

Jointly, this implies a simple strategy for determining the minimum connectivity $\muT$ below the first gap of PhCs:
\begin{enumerate}[itemsep=0ex,parsep=0.5ex,left=1.5ex]
    \item pick a longitudinal solution $\tilde{\nv}^{\text{L}}$ with connectivity $\muL$ and define $\muLT = \muL + 2$,
    \item find all apolar solutions $\{\nv^{\text{L+T}}\}$ with connectivity $\muLT$,\label{step:apolar_solutions}
    \item if any $\{\nv^{\text{L+T}}\} - \tilde{\nv}^{\text{L}}$ are valid, \ie respect \cref{eq:gamma_constraints,eq:regularity_constraint}, they represent physically realizable transverse solutions $\{\nv^{\text{T}}\}$ with connectivity $\muT=\muLT - \muL$; if not, increment $\muLT$ and return to step~\ref{step:apolar_solutions}.
\end{enumerate}
The Hilbert basis $\{\mathbf{h}_i\}$ allows a highly efficient and exhaustive computation of the finite set of solutions in step~\ref{step:apolar_solutions} without the combinatorial challenges that a non-conical basis would present (\suppsec~\ref{supp-sec:compute_connectivtity_solutions}).
Non-minimal connectivity solutions can be obtained by simply continuing the iteration procedure (\suppsec~\ref{supp-sec:nonminimal_connectivity}).

\begin{figure*}[!htb]
	\centering
	\includegraphics[scale=1]{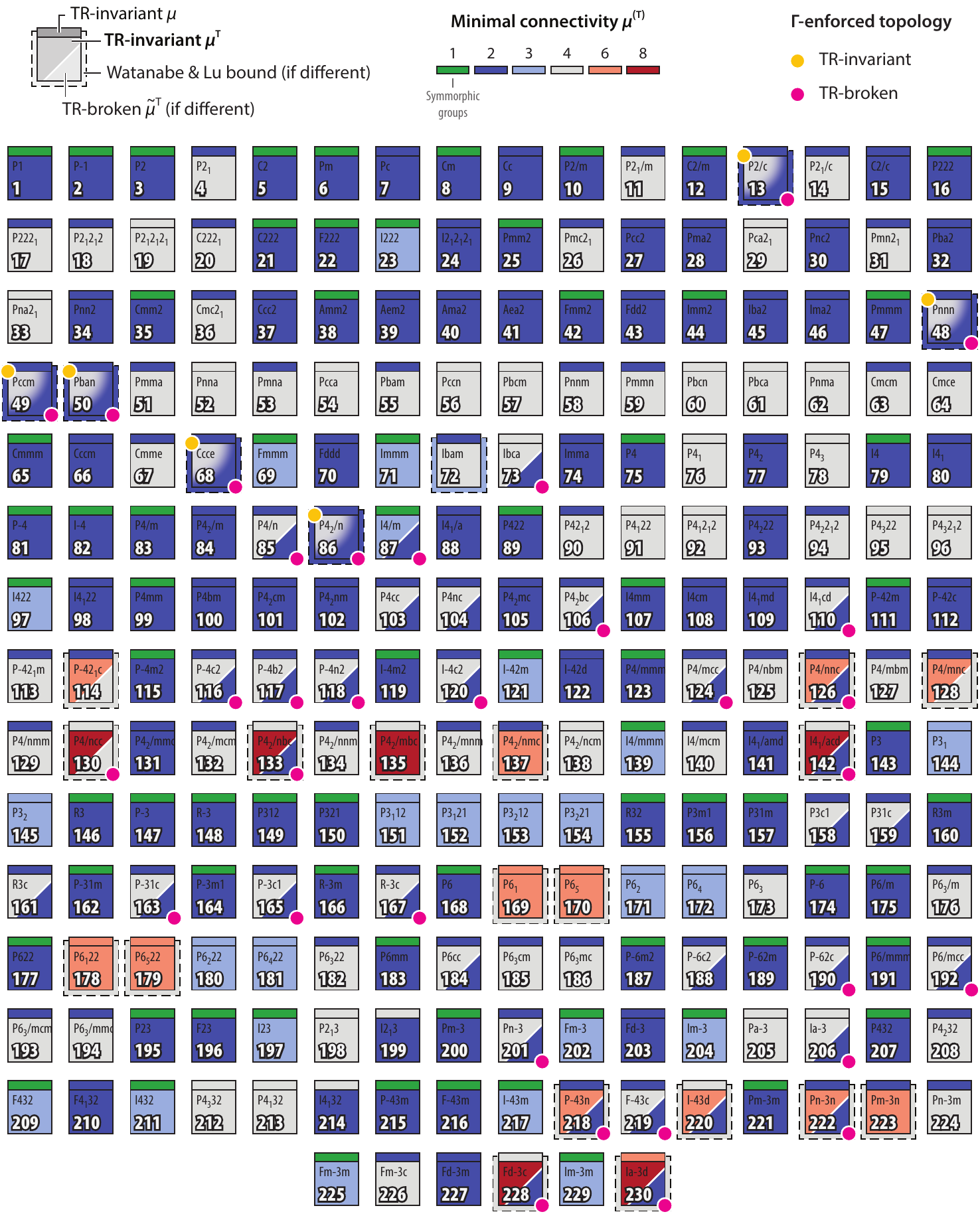}
	\caption{%
	    \textbf{Minimum photonic band connectivity and $\mathsf{\Gamma}$-enforced topology.}
	    The TR-invariant minimum transverse band connectivity below the first gap, $\muT$, is given for each space group (labeled squares).
        Watanabe and Lu's~\cite{Watanabe:2018} lower bounds are shown (dashed boxes) when our explicit solutions exceed them.
        The corresponding TR-broken connectivities $\muTbtr$ are shown (lower triangular cut-outs) when they differ from their TR-invariant counterparts.
        Minimum-connectivity solutions with $\mathsf{\Gamma}$-enforced topology are highlighted by circular markers (in yellow, with TR; in pink, without TR).
        TR-invariant minimum regular connectivities $\mu$ are shown as context (narrow rectangles).
        Compatibility relations allow $\muT=2$ solutions in space groups 13, 48\,--\,50, 68 and 86 (lower-diagonal box-shading) that, however, are topologically obstructed (\ie incompatible with a gap) due to $\mathsf{\Gamma}$-enforced non-gaps which increases $\muT$ to $4$ (upper-diagonal box-shading).%
		\label{fig:2}
        }
\end{figure*}

\section{Results}\label{sec:results}

\subsection{Minimum photonic band connectivity}\label{subsec:connectivity}
We applied our framework to compute the minimum-connectivity transverse solutions for each of the 230 space groups~%
    \footnote{The 11 enantiomorphic pairs (chiral space groups related by a mirror)~\cite{Nespolo:2018} must have identical global properties. 
    More precisely then, $\muT$ is a property of the 219 affine space group types.}%
, with and without TR symmetry~%
    \footnote{Incorporation of TR symmetry requires only our swapping out EBRs and irreps for their TR-symmetric counterparts, \ie physically real EBRs~\cite{Bradlyn:2017} and irreps (coreps)~\cite{Herring:1937, BradleyCracknell:1972}.}.
As one of the central results of this work, \cref{fig:2} summarizes the associated minimum transverse connectivities $\muT$ versus space group.
In addition to the mininum transverse connectivity, we also indicate the minimum regular connectivity $\mu$ (with TR) which applies to all bands above the first gap.
As previously noted by \citet{Watanabe:2018}, $\muT$ is neither smaller nor larger than $\mu$ in general:
\eg in all symmorphic space groups, the regular connectivity is $\mu = 1$ but the transverse connectivity $\muT$ is larger, equaling either 2 or 3 (being at least 2 due to the double degeneracy at $\omega=0$).
Conversely, the cubic space groups 199 (I2\textsubscript{1}3) and 214 (I4\textsubscript{1}32; single gyroid) have $\muT = 2$ \emph{smaller} than $\mu = 4$.
Ref.~\citenum{Watanabe:2018} established the existence of $\muT=2$ solutions for 104 space groups and obtained lower bounds on $\muT>2$ for the remaining groups (with TR) by manually deriving compatibility-respecting solutions for 38 key groups in combination with translationengleiche~\cite{ITA:6} sub- and supergroup relations.
By evaluating all solutions explicitly, we find 19 exceptions---namely, space groups 72, 114, 126, 128, 130, 133, 135, 137, 142, 169, 170, 178, 179, 218, 220, 222, 223, 228, and 230 (\cref{fig:2}, dashed boxes)---that exceed these lower bounds (as we discuss later, accounting for topology reveals additional exceptions).
The exceptions are all nonsymmorphic space groups and associate with the presence of additional screw or glide axes---or, rarely, with inversion---relative to the considered key subgroup~%
    \footnote{As an example, space group 85 (P4/n) is a key group for space group 126 (P4/nnc): while P4/n is generated by inversion $-1$ and a four-fold screw $\{4_{001}^+ | \tfrac{1}{2}00\}$ and has $\muT = 2$, P4/nnc contains an additional 2-fold screw (or, equivalently, a glide) axis among its generators, \eg $\{2_{100} | \tfrac{1}{2}00\}$, and has $\muT = 4$.}.
Generally, we observe that space groups with $\muT > 2$ are either nonsymmorphic or body- or face-centered, consistently with Ref.~\citenum{Watanabe:2018}.
The exact impact of nonsymmorphic symmetry is varied and detail-sensitive:
as an example, the nonsymmorphic tetragonal space groups 112 (P$\overline{\text{4}}$2c), 113 (P$\overline{\text{4}}$2\textsubscript{1}m), and 114 (P$\overline{\text{4}}$2\textsubscript{1}c) have $\muT$ equal to 2, 4, and 6, respectively---despite having identical point group symmetry ($\overline{\text{4}}$2m or $D_{2d}$) and screws and glides that differ only in their translation parts.

We compared our results for $\muT$ with a recent high-throughput computational search for PhC band gaps by \citet{CersonskyGlotzer:2021}.
In this search, 103 space groups were identified by explicit examples as capable of hosting complete PhC gaps at a dielectric constrast below $16$.
In each such case, the computed number of bands below the first gap in Ref.~\citenum{CersonskyGlotzer:2021} are consistent with the $\muT$ reported here (\ie equal to or higher, with equality attained in 31 space groups)~%
    \footnote{We note a single disagreement with Ref.~\citenum{CersonskyGlotzer:2021}, who report a PhC with a \SI{0.1}{\percent} gap between bands 2 and 3 in space group 224, inconsistently with $\muT=4$ found here.
    This gap appears to be spurious and is likely due to numerical symmetry-breaking.}.

\begin{figure*}[!htb]
	\centering
	\includegraphics[scale=1]{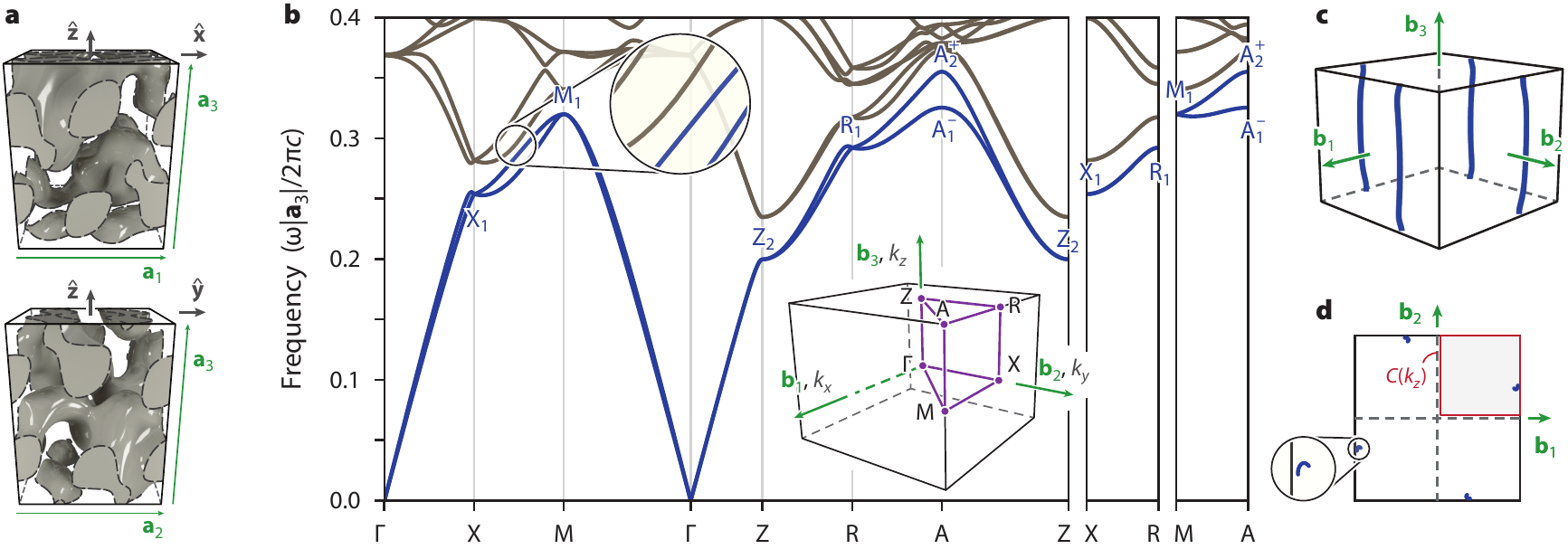}
	\caption{%
	    \textbf{Photonic topological non-gap in space group 86 (P4\textsubscript{2}/n).}
        \captionbullet{a}~PhC unit cell and
        \captionbullet{b}~associated band structure (along indicated BZ path) that realizes a minimum-connectivity ($\mu^{\text{T}}=2$) $\mathsf{\Gamma}$-enforced topological solution in space group 86 ($\mu^{\text{T}}=2$) $\mathsf{\Gamma}$-enforced topological solution in space group 68 (unit cell parameterization in Supplemental Table~\ref{tab:lattice_86}; $\varepsilonup=16$ material in gray, embedded in vacuum with a filling fraction of \SI{35}{\%}; $\mathbf{a}_i$ and $\mathbf{b}_i$, direct and reciprocal lattice vectors)
	    \captionbullet{c}~Four nodal lines connect bands 2 and 3, running along $k_z$ at generic $(k_x,k_y)$.
        \captionbullet{d}~Projection of nodal lines to the $(k_x,k_y)$ plane and definition of a fixed-$k_z$ loop $C(k_z)$ which enclosing a quadrant of the $(k_x,k_y)$ plane and a nodal line.
        Bands 1, 2, 3, and 4 do not touch on $C(k_z)$: the Berry phase of band 2 and 3 (bands 1 and 4) around $C(k_z)$ is $\pi$ ($0$), protecting an odd multiple of nodal lines in the interior of $C(k_z)$.
        \label{fig:3}
	}
\end{figure*}

By breaking TR-symmetry, \eg via an external magnetic field, irreps that otherwise stick together due to TR are split (\ie complex or pseudoreal irreps~\cite{Herring:1937}).
In 36 space groups this leads to a reduction of the TR-broken transverse connectivity $\muTbtr$ relative to its TR-invariant value $\muT$ (\cref{fig:2}, triangular cut-outs).
The reduction often corresponds to the splitting of a Hilbert basis vector in the TR-invariant solution into two TR-broken components that were otherwise held together by a (self-)conjugate irrep pair; \eg in space group 161 (R3c), splitting the physically real 4D irrep $T_3T_3$ to two pseudoreal 2D $T_3$ irreps lowers $\muT=4$ to $\muTbtr=2$.
In these cases, applying a TR-breaking perturbation to a TR-invariant minimum-connectivity solution will necessarily lower the connectivity from $\muT$ to $\muTbtr$.
Interestingly, several $\muTbtr<\muT$ solutions cannot be obtained in this perturbative fashion.
For instance, the TR-invariant $\muT = 4$ solutions of space group 73 (Ibca) require the $X$-point symmetry $X_1^+ + X_2^+ + X_3^+ + X_4^+$ or $X_1^- + X_2^- + X_3^- + X_4^-$ while the TR-broken $\muTbtr = 2$ solution requires $X_4^+ + X_4^-$ symmetry; the latter cannot be decomposed from the former, and hence also not by perturbatively breaking TR in it.

As in the example above, a given connectivity can usually be realized by multiple symmetry vectors $\{\nv^{\text{T}}\}$.
The associated \emph{regular} symmetry content is definite and physical---\ie $\{\nv_\Gamma^{\text{T}}\} - \nv^{\text{2T}}_\Gamma$ and $\{\nv_{\kv\neq\Gamma}^{\text{T}}\}$ are physical quantities---and we exhaustively enumerate every minimum-connectivity symmetry vector in \suppsec{}s~\ref{supp-sec:connectivity-invariant-tr} and \ref{supp-sec:connectivity-broken-tr}.
In the following, we introduce a method to evaluate the transverse solutions' topology despite their singular $\Gamma$-point symmetry.
Surprisingly, we find that topology can constrain $\muT$ beyond the requirements imposed by compatibility relations.

\subsection{Topology of singular transverse bands}\label{subsec:topology}
Any regular symmetry vector can be mapped to topological indices $(\nu_1, \ldots, \nu_{\lambda_{d^{\text{BS}}}})$ in the symmetry indicator group $X_{\text{BS}} = \mathbb{Z}_{\lambda_1}\times \ldots\times\mathbb{Z}_{\lambda_{d^{\text{BS}}}}$ with $\mathbb{Z}_{\lambda_i} \equiv \{0, 1, \ldots, \lambda_i-1\}$ (see \suppsec~\ref{supp-sec:band-structures}--\ref{supp-sec:topology_summary})~\cite{Po:2017}.
For brevity, we omit trivial factor groups, \ie write $\mathbb{Z}_{\lambda_i}\times\mathbb{Z}_1$ as $\mathbb{Z}_{\lambda_i}$ and $\mathbb{Z}_1 \times \ldots \times \mathbb{Z}_1$ as $\mathbb{Z}_1$.
Space groups with symmetry-identifiable topology have indicator groups $X_{\text{BS}}\neq \mathbb{Z}_1$ (equivalently, $\nu_i= 0$ denotes a trivial index): with TR symmetry, there are 53 such space groups, all corresponding to topological nodal features~\cite{Song:2018}.
Denoting by $\mathbf{B}$ the column-wise matrix-concatenation of EBR vectors, and by $\mathbf{B} = \mathbf{S}\boldsymbol{\Lambda}\mathbf{T}$ its associated Smith normal decomposition, the topological indices of a regular symmetry vector $\nv\in\{\text{BS}\}$ are~\cite{Tang:2019}:
\begin{equation}\label{eq:topology}
    \nu_i = \mathbf{S}^{-1}_{i,\star}\nv\ \mathrm{mod}\ \lambda_i,
\end{equation}
with $\mathbf{S}^{-1}_{i,\star}$ denoting the $i$th row of $\mathbf{S}^{-1}$ and $\lambda_i$ the $i$th diagonal element of $\boldsymbol{\Lambda}$  (corresponding to the indicator group's $\mathbb{Z}_{\lambda_i}$ term).

For photonic bands below the first gap, a similar approach is not workable since the associated symmetry vector $\nv^{\text{T}}$ may be singular (\ie $\nv^{\text{T}}$ may not belong to \BS).
Instead, in the spirit of $K$ theory, we define indices for $\nv^{\text{T}}$ by considering the \emph{difference} of the apolar ($\nu_i^{\text{L}+\text{T}}$) and longitudinal ($\nu_i^{\text{L}}$) indices, whose symmetry vectors are regular:
\begin{equation}\label{eq:transverse_topology}
    \nu_i^{\text{T}}
    =
    \big(\nu_i^{\text{L}+\text{T}} - \nu_i^{\text{L}}\big)\ \mathrm{mod}\ \lambda_i.
\end{equation}
This is a key result of our work: it enables direct and full application of symmetry-based diagnosis for band topology to all 3D PhCs, despite the $\Gamma$-point singularity at $\omega = 0$.
Crucially, while the auxiliary indices $\nu_i^{\text{L}+\text{T}}$ and $\nu_i^{\text{L}}$ are not unique---\cref{eq:gamma_constraints} leaves substantial freedom of choice for the auxiliary symmetry vectors---their difference is (\suppsec~\ref{supp-sec:proof_topological_relation}).

\subsection{Photonic topological non-gaps}\label{subsec:nongaps}
Using \cref{eq:transverse_topology}, we evaluated the symmetry-indicated topology of every minimum-connectivity transverse solution (\suppsec{}s~\ref{supp-sec:connectivity-invariant-tr} and \ref{supp-sec:connectivity-broken-tr}).
In doing so, we discover six centrosymmetric and nonsymmorphic space groups---13 (P2/c), 48 (Pnnn), 49 (Pccm), 50 (Pban), 68 (Ccce), and 86 (P4\textsubscript{2}/n)---whose minimum-connectivity ($\muT=2$) transverse solutions are \emph{all} topologically nontrivial (\cref{fig:2}, yellow markers).
If a gap exists between bands two and three in the high-symmetry band structure of these space groups, the resulting gap is guaranteed to be topologically nontrivial---\ie nontrivial topology is implied by band connectivity alone.
While such connectivity-implied nontriviality is reminiscent of filling-enforced topology in the electronic context~\cite{Po:2016, Wang:2020}, we stress that its appearance in the photonic context is intrinsically different since it is inseparable from the $\Gamma$-point singularity---accordingly, we will refer to it as ``$\Gamma$-enforced topology.''
In contrast, the minimum-connectivity solutions of TR-invariant regular bosons (\eg photonic bands above the first gap) do not display $\Gamma$-enforced topology (\suppsec~\ref{supp-sec:fe-regular}).

\begin{figure*}[!htb]
	\centering
	\includegraphics[scale=1]{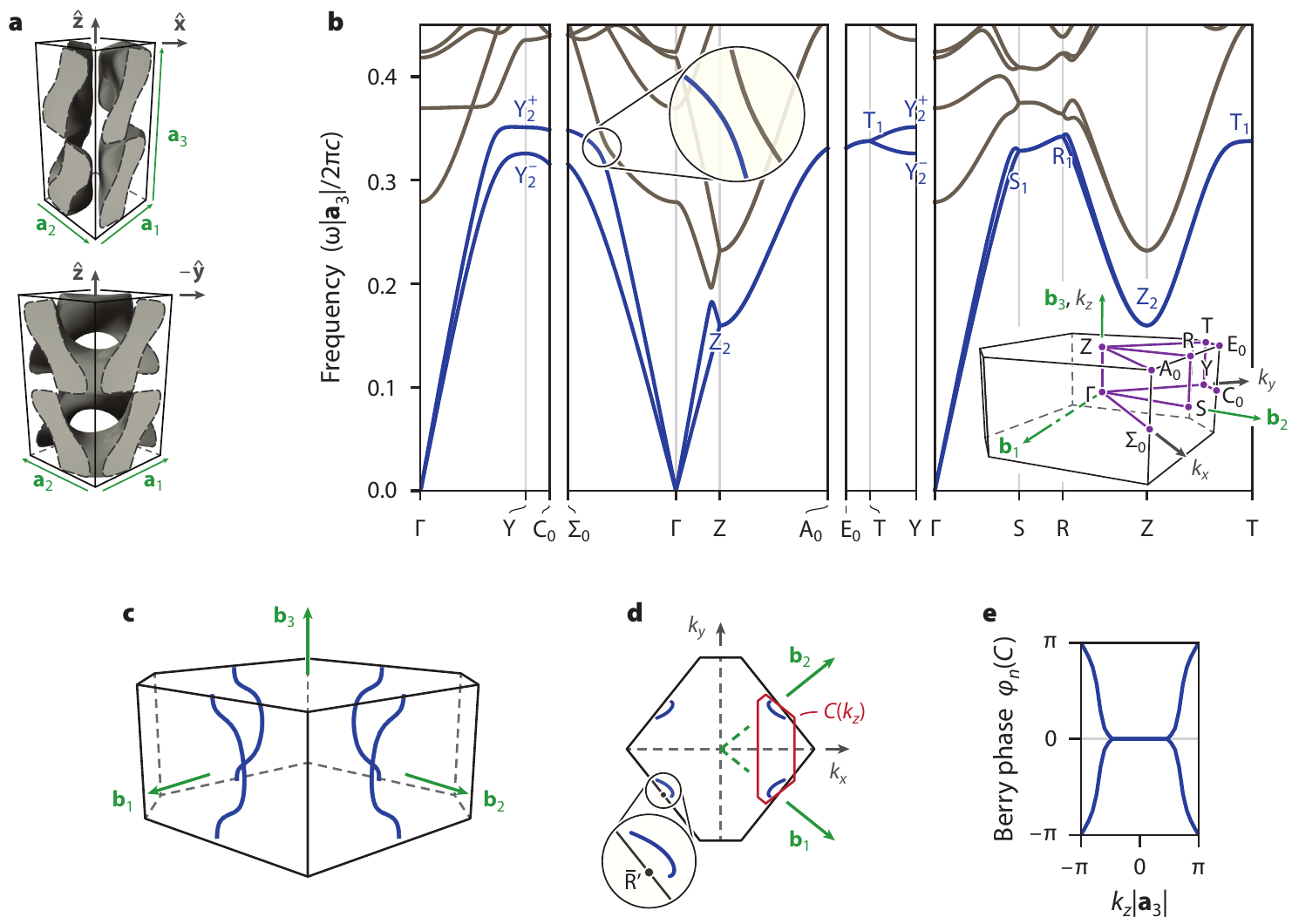}
	\caption{%
	    \textbf{Photonic topological non-gap in space group 68 (Ccce) with $\mathbb{Z}_2$ monopole charge.}
        \captionbullet{a}~PhC unit cell and
        \captionbullet{b}~associated band structure (along indicated BZ path) that realizes a minimum-connectivity ($\mu^{\text{T}}=2$) $\mathsf{\Gamma}$-enforced topological solution in space group 68 (unit cell parameterization in Supplemental Table~\ref{tab:lattice_68}; $\varepsilonup=16$ material in gray, embedded in vacuum with a filling fraction of \SI{35}{\%}; $\mathbf{a}_i$ and $\mathbf{b}_i$, direct and reciprocal primitive lattice vectors).
	    \captionbullet{c}~Nodal lines connect bands 2 and 3 at generic momenta. 
        \captionbullet{d}~Projection of nodal lines to the $(k_x,k_y)$ plane and definition of a fixed-$k_z$ loop $C(k_z)$ that encloses a pair of nodal lines.
        \captionbullet{e}~The non-Abelian Berry phases around $C(k_z)$ of bands 1 and 2 wind relatively when $k_z$ ranges over its domain.
        The nodal lines are consequently protected by a $\mathbb{Z}_2$ monopole charge (in addition to a $\pi$-Berry phase).
		\label{fig:4}
	}
\end{figure*}

Notably, all TR-invariant symmetry-identifiable bosonic topology is associated with bulk nodal features: specifically, with Weyl points and nodal lines for centro- and noncentrosymmetic space groups, respectively~\cite{Song:2018}.
Intriguingly, for the six space groups with $\Gamma$-enforced topology, observing a $\muT = 2$ gap along high-symmetry paths in the BZ thus necessarily implies the existence of gap-closing nodal lines between bands 2 and 3 at generic momenta.
We refer to this as a topological ``non-gap'': a high-symmetry gap that implies a gap-closing.
Non-gaps demonstrate a new, topological constraint on photonic band connectivity beyond compatibility constraints.
With this constraint accounted for, $\muT$ is increased from 2 to 4 for space groups 13, 48--50, 68, and 86 (\suppsec~\ref{supp-sec:nonminimal_connectivity-fe}).

We next demonstrate a concrete PhC with a topological non-gap in space groups 86 (generated by inversion $-1$ and screws $\{2_{001}|\tfrac{1}{2} \tfrac{1}{2} 0\}$ and $\{4_{001}^+| 0\tfrac{1}{2} \tfrac{1}{2}\}$).
The $\muT = 2$ compatibility-respecting solutions of space group 86 are $\nv^{\text{T}} = [(\smallsquare)^{2\text{T}}, A_1^{\mp} + A_2^{\pm}, M_1, Z_2, R_1, X_1]$---with $(\smallsquare)^{2\text{T}}$ indicating the singular $\Gamma$-point symmetry---with nontrivial index $\nu_1^{\text{T}} = 1 \in \mathbb{Z}_2$.
We performed a random search of PhCs spanned by symmetry-respecting Fourier-sum isosurfaces (\suppsec~\ref{supp-sec:unitcell}), checking the symmetry vectors of their lowest two bands against $\nv^{\text{T}}$ to identify a realization~%
    \footnote{Our calculations use the MPB frequency-domain solver~\cite{Johnson:2001}.}.
\Cref{fig:3}ab shows one such PhC realization: its associated band structure shows a $\muT = 2$ connectivity along the high-symmetry BZ paths.
For regular TR-invariant bosonic bands in space group 86, a $\nu_1 = 1$ index protects a $\pi$-Berry phase in each quadrant of any $k_z$-slice of the BZ (stabilized by the combination of TR- and inversion symmetry), which in turn requires the existence of $4\Mod{8}$ nodal lines running along $k_z$~\cite{Song:2018}.
Consistently with this, our PhC hosts 4 nodal lines between bands 2 and 3, each protected by a $\pi$-Berry phase in band 2 (\cref{fig:3}cd).

Unlike space group 86, the nodal lines of space group 48--50 and 68 are protected not only by a $\pi$-Berry phase but also by a $\mathbb{Z}_2$ monopole charge (space group 13 may exhibit either type of protection)~\cite{Song:2018}.
\Cref{fig:4} shows a PhC realization in space group 68 (generated by inversion $-1$, screws $\{2_{001}|\tfrac{1}{2} 0 0\}$ and $\{2_{010}|0 0 \tfrac{1}{2}\}$, and a $C$-centering translation $\{1 | \tfrac{1}{2} \tfrac{1}{2} 0\}$).
The PhC realizes the sole $\muT=2$ compatibility-respecting solution, with symmetry vector $\nv^{\text{T}} = [(\smallsquare)^{2\text{T}}, T_1, Y_2^+ + Y_2^-, Z_2, R_1, S_1]$ and index $\nu_1^{\text{T}} = 1\in\mathbb{Z}_2$.
The high-symmetry band structure thus exhibits a $\muT=2$ gap (\cref{fig:4}b)---but there are nodal lines closing the gap at generic momenta (\cref{fig:4}c).
For any loop enclosing a single nodal line, the summed Berry phase of bands 1 and 2 is $\pi$, as before.
More interestingly, for a loop $C(k_z)$ enclosing a pair of nodal lines (\cref{fig:4}d), bands 1 and 2 (and 3 and 4) necessarily touch as $k_z$ is wound over its domain~%
    \footnote{The loop $C(k_z)$ must intersect the $\kv$-line $A = \alpha\mathbf{b}_1 + \alpha\mathbf{b}_2 + \tfrac{1}{2}\mathbf{b}_3$ (primitive reciprocal lattice vectors $\mathbf{b}_i$) whose only irrep is 2D.}.
Computing the non-Abelian Berry phases~\cite{WilczekZee:1984, Soluyanov:2012, Vanderbilt:2018} of bands 1 and 2 over $C(k_z)$ while winding $k_z$, we observe a relatively winding Berry phase spectrum signaling an enclosed $\mathbb{Z}_2$ monopole charge (\cref{fig:4}e)~\cite{Yu:2011, Ahn:2018}, consistently with the general predictions from topological band theory~\cite{Song:2018, Note999}.
These $\mathbb{Z}_2$-charged nodal lines are protected against gapping by any TR- and inversion-preserving perturbations, just as the more conventional $\pi$-Berry phase nodal lines, but additionally can only be created and annihilated in pairs~\cite{Fang:2015, Kim:2015, Ahn:2018}.%
    \footnotetext[999]{There is an inconsequential cosmetic difference between the nodal lines in \cref{fig:4}c and those predicted in Ref.~\citenum{Song:2018}: we observe paired nodal lines rather than loops; the former can be obtained from the latter by merging across the $k_z=0$ plane}
The impact of the distinct topological protections in space group 86 and 68 could be detected experimentally since surface states extending from the projections of the nodal lines, \eg at $[001]$ facets, are single-helicoid for $\pi$-Berry phase lines but double-helicoid for $\mathbb{Z}_2$-charged lines~\cite{Fang:2016, ChengLu:2020}.

We also searched for $\Gamma$-enforced topology in the TR-broken setting, finding 32 space groups (\cref{fig:2}, pink markers).
Since nontrivial TR-broken topology also includes gapped phases, however, this $\Gamma$-enforced topology does not necessarily correspond to topological non-gaps.
A prominent mechanism for breaking TR for photons involves applying a uniform magnetic field to magneto-optic PhC~\cite{Wang:2009}.
Among the identified candidates, only a space groups 13, 85, 86, and 87 have symmetries compatible with a uniform magnetic field.
Of these, only space groups 85 and 87 require TR-breaking in order to exhibit $\Gamma$-enforced topology.
By analyzing their solutions' symmetry, we find that the TR-broken minimum-connectivity ($\muTbtr=2$) solutions of either case \emph{cannot} be obtained by perturbatively applying TR-breaking to their TR-invariant minimum-connectivity solutions.
Instead, they require either large-amplitude TR-breaking or fine-tuned accidental degeneracies (\suppsec~\ref{supp-sec:trbreaking-fe}).

Our discovery of topological non-gaps is an interesting counter-example to the usual working assumption that a gap along the high-symmetry lines and edges of the BZ implies a full gap. 
Here, instead, the converse is guaranteed.
While exceptions to the rule are well-known~\cite{Harrison:2007, Craster:2012}, it is true that \emph{most} band extrema occur at the BZ edge or along high-symmetry lines~\cite{Harrison:2007, Maurin:2018}.
Our findings are consistent with this perspective since nodal lines are linear band degeneracies, not band extrema.

\begin{figure*}[!htb]
    \centering
    \includegraphics{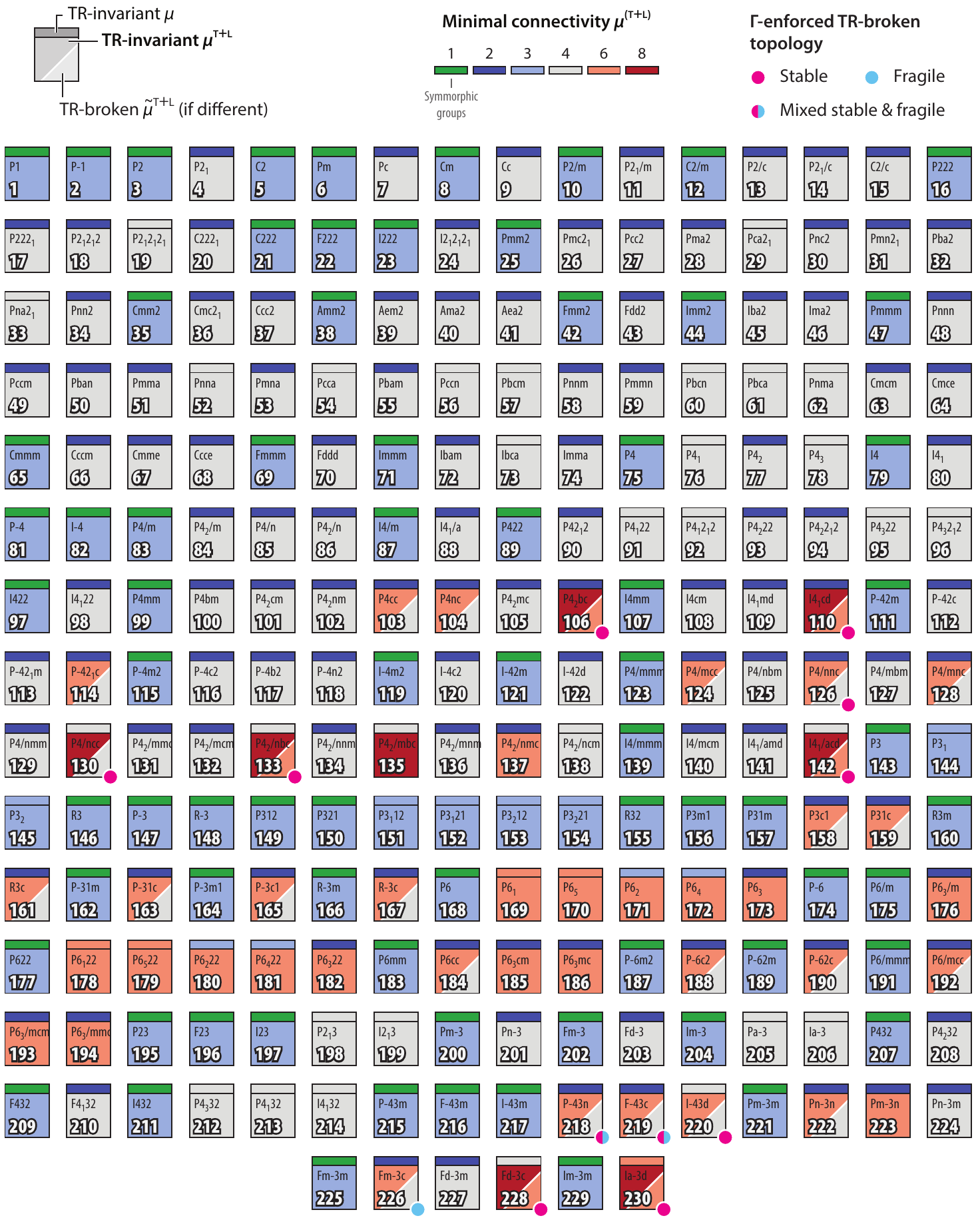}
    \caption{%
	    \textbf{Minimum phononic and acoustic wave band connectivity and $\mathsf{\Gamma}$-enforced topology.}
        Conventions as in \cref{fig:2}.
        \label{fig:phononic-connectivity-table}
    }
\end{figure*}


\section{Discussion}

We have here explicitly constructed the photonic minimum connectivity below the first gap by requiring consistency with compatibility relations and pinned symmetry content at $\omega=0$.
A natural question is whether every such connectivity can be realized with available optical materials.
Space group 230 (Ia$\overline{3}$d), which \eg hosts the double gyroid structure~\cite{Lu:2013, Lu:2015}, provides an interesting counter-example:
while we predict the possibility of a $\muT=8$ gap (\cref{fig:2}), to the best of our knowledge, all previously observed gaps exhibit a higher connectivity~\cite{Lu:2013, CersonskyGlotzer:2021}.
This apparent discrepancy is caused by what we will call a ``dielectric obstruction''---a difficulty in bringing together the band symmetries required by a given solution due to material constraints.
In more detail, every $\muT = 8$ solution in Ia$\overline{3}$d requires a $\Gamma_1^+$ or $\Gamma_1^-$ irrep, or both (\suppsec~\ref{supp-sec:connectivity-invariant-tr}).
These irreps, however, occur only in the very high-lying bands of PhCs in Ia$\overline{3}$d.
To understand this, we consider the irreps of the empty-lattice structure (\ie constant $\varepsilonup$): there, no $\Gamma_1^\pm$ features in the 38 lowest-frequency modes at $\Gamma$ (\suppsec~\ref{supp-sec:sg230}).
To gradually transform the empty-lattice to a hypothetical $\muT=8$ PhC then requires at least 38 band-inversions at $\Gamma$---necessitating an extremely large dielectric contrast, likely outside the attainable range.
Precisely where such dielectric obstructions arise is an interesting question, with implications for photonic topology and band-engineering in general.

Our construction relies explicitly on the use of a set of auxiliary apolar and longitudinal modes to circumvent the symmetry singularity at $\omega=0$.
This need arises because the transversality condition requires the discarding of one third of all solutions to the Maxwell equations---the longitudinal modes---fracturing the symmetry content at $\omega=0$.
Unlike photons, phonons and acoustic waves allow both transverse and longitudinal polarizations; like photons, they intersect $\omega=0$ at $\Gamma$ but in a triply degenerate fashion~\cite{Kittel:1987}.
Their symmetry content at $\omega=0$ is consequently not singular but is still pinned by their long-wavelength plane-wave-like behavior.
Phonons and acoustic waves are therefore subject to the apolar $\nv_\Gamma^{\text{1L+2T}}$ constraint in \cref{eq:gamma_constraints}: the connected phononic solutions below the first gap are simply all the non-negative combinations of Hilbert vectors that each contribute to fulfilling this constraint.
The minimum phononic connectivity below the first gap $\muLT$ (\cref{fig:phononic-connectivity-table}) can consequently be computed immediately with the tools already developed.
The corresponding solutions are regular and their (stable) topology can be evaluated directly from \cref{eq:topology}---in fact, regularity also enables evaluation of fragile topology~\cite{Po:2018} (by checking non-negative integer expansion feasibility in the EBR basis; see \suppsec~\ref{supp-sec:topology_summary}).
Doing so for every space group, we find no $\Gamma$-enforced topology in the TR-invariant setting but 9 space groups with TR-broken $\Gamma$-enforced stable topology and 3 with fragile or mixed stable--fragile topology (\suppsec~\ref{supp-sec:phononic-fetop-tables}).
Such $\Gamma$-enforced topological phonons, or magnetic space group analogues~\cite{Watanabe:2018b, Elcoro:2020b}, may be realizable in ferroelectric materials.
A related application is to the connectivity of bulk longitudinal plasmons of metals (associated with zeros of the longitudinal nonlocal dielectric function~\cite{Pines:1952}) or to certain recently proposed perfect-metal metamaterials~\cite{Chen:2018} (governed by the quasistatic Poisson equation), which, if wholly decoupled from transverse fields, are subject only to the $\nv_\Gamma^{\text{1L}}$ constraint in \cref{eq:gamma_constraints}.

In this work, we provide explicit constructions of all minimum-connectivity solutions below the first gap in PhCs.
Our approach exploits the existence of a Hilbert basis $\{\hv_i\}$, consisting of all minimally connected bands, to very efficiently and exhaustively solve constrained connectivity problems.
To apply this technique to the singular $\omega=0$-connected transverse PhC bands, we introduce two auxiliary, regular problems associated with bands of apolar and longitudinal polarizations, whose difference correspond to the physical transverse bands.
Leveraging this decomposition further, we introduce a definition for the symmetry-identifiable topological indices of the singular PhC bands below the first gap in \cref{eq:transverse_topology}, thereby overcoming the key technical barrier to application of symmetry-based diagnosis of band topology to 3D PhCs~\cite{Bradlyn:2019}.
By exhaustive computation of the topology of the minimum-connectivity solutions in every space group, we discover the existence of photonic topological non-gaps---gaps in the high-symmetry band structure whose existence imply necessary band closings, in the form of nodal lines, in the interior of the BZ---in space groups 13, 48--50, 68, and 86, providing proof-of-concept examples for the latter two.

The singular nature of the $\omega=0$ bands is a uniquely photonic feature, arising as a direct result of the transverse polarization of photons.
Beyond the problems considered here, the singularity's manifestations are at the core of several problems in electromagnetic theory, including \eg unique photonic considerations for $\mathbf{k}\cdot{\mathbf{p}}$ models~\cite{Sipe:2000} and long-standing obstacles to the construction of exponentially localized Wannier functions in 3D PhCs~\cite{Wolff:2013}.
We expect the ideas introduced here may be useful also in these directions.

\FloatBarrier
\section*{Acknowledgements}
We thank Ling Lu, Steven G.\ Johnson, and Robert-Jan Slager for stimulating discussions.
This research is based upon work supported in part by the Air Force Office of Scientific Research under the award number FA9550-20-1-0115,
the US Office of Naval Research (ONR) Multidisciplinary University Research Initiative (MURI) grant N00014-20-1-2325 on Robust Photonic Materials with High-Order Topological Protection
and the U.S.\ Army Research Office through the Institute for Soldier Nanotechnologies at MIT under Collaborative Agreement Number W911NF-18-2-0048.
The work of H.C.P.\ is partly supported by a Pappalardo Fellowship at MIT.
The MIT SuperCloud and Lincoln Laboratory Supercomputing Center provided computing resources that contributed to the results reported in this work.

\section*{Code availability}
The software tools developed in this work are made available as open-source software implemented in the Julia programming language~\cite{Crystalline.jl, SymmetryBases.jl, PhotonicBandConnectivity.jl}.
Functionality to compute general symmetry eigenvalues of PhC eigenstates was added to the MPB frequency-domain solver~\cite{Johnson:2001} as a part of this work.

\FloatBarrier
\bibliographystyle{apsrev4-2-longbib}
\bibliography{references}

\FloatBarrier
\clearpage
\foreach \x in {1,...,344} 
{%
\clearpage
\includepdf[noautoscale=true,pages={\x}]{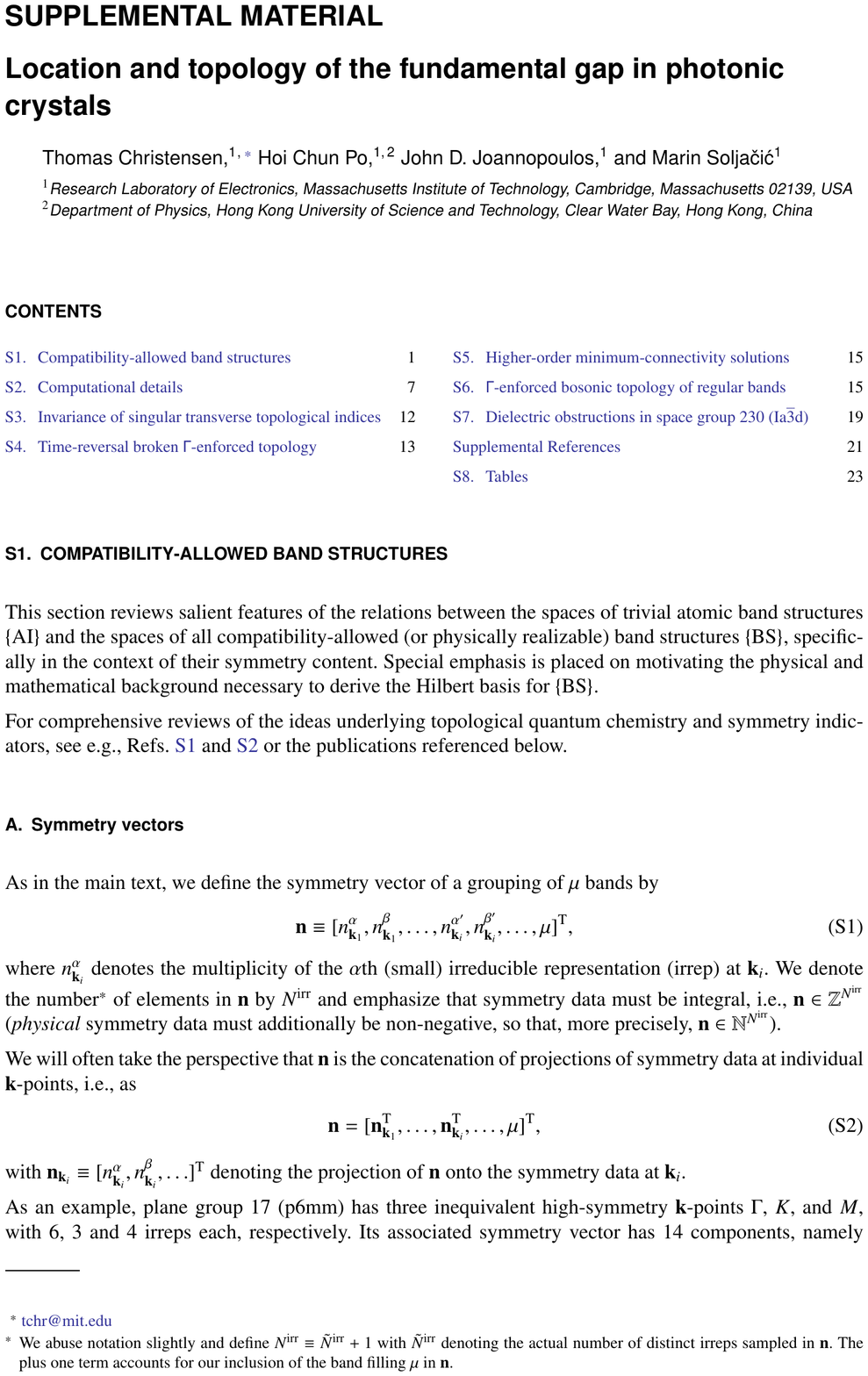} 
}

\end{document}